\documentstyle[epsfig,twocolumn,prb,aps]{revtex}
\newcommand{\figwidth}{6.0cm} 
\newcommand{\figheight}{8.6cm} 

\begin{document}

\draft 
\twocolumn[\hsize\textwidth\columnwidth\hsize\csname @twocolumnfalse\endcsname

\title{
  Phases of the one-dimensional Bose-Hubbard model
} 

\author{T.~D.~K\"uhner and H.~Monien} 

\address{
  Physikalisches Institut der Universit\"at Bonn,
  D-53115 Bonn, 
  Germany
}

\date{\today}  \maketitle 

\widetext

\begin{abstract}
  \noindent
  The zero-temperature phase diagram of the one-dimensional Bose-Hubbard
  model with nearest-neighbor interaction is investigated using the
  Density-Matrix Renormalization Group.  Recently {\it normal} phases
  without long-range order have been conjectured between the charge
  density wave phase and the superfluid phase in one-dimensional bosonic
  systems without disorder. Our calculations demonstrate that there is
  no intermediate phase in the one-dimensional Bose-Hubbard model but a
  simultaneous vanishing of crystalline order and appearance of
  superfluid order.  The complete phase diagrams with and without
  nearest-neighbor interaction are obtained.  Both phase diagrams show
  reentrance from the superfluid phase to the insulator phase.
\end{abstract}

]

\narrowtext

Quantum phase transitions in strongly correlated systems have
attracted a lot of interest in recent years. Usually the basic
particles are electrons, but in some interesting cases the relevant
particles are not fermions but bosons.  Examples of experimental
systems with superfluid and insulating phases are Cooper pairs in thin
granular superconducting films \cite{Orr.Haviland.Jaeger} and cooper
pairs or fluxes in Josephson junction arrays\cite{Oudenaarden:1996}.
While in dimensions greater than one the existence of supersolids
\cite{MeiselBatrouni}, i.e. phases with simultaneous superfluid and
crystalline order, has been established in theoretical work, the
situation in one dimension is less clear.  Recently normal phases that
are neither crystalline nor superfluid have been found in a
one-dimensional model of Josephson junction arrays \cite{Baltin} in
the region where supersolids are found in higher dimensions.  In this
paper we will verify whether supersolids or normal phases exist in the
more general Bose-Hubbard model in one dimension.

The Bose-Hubbard model contains the basic physics of interacting bosons 
on a lattice. It is a minimal bosonic many-particle model that cannot
be reduced to a single particle model.  The bosons have repulsive 
interactions, and they can gain energy by hopping to neighboring sites on
the lattice. The Hamiltonian with on-site and nearest-neighbor 
interactions is
\begin{eqnarray}
  \label{eq.BH2}
  \nonumber
  H_{BH}  =&  - t \sum_i ( b^{\dagger}_{i} b^{\phantom{\dagger}}_{i+1}
  +  b^{\phantom{\dagger}}_{i} b_{i+1}^{\dagger} ) - \sum_{i} \mu n_{i} \\
  &  + U \sum_{i} n_{i} ( n_{i} - 1 )/2
  + V \sum_{i}  n_{i} n_{i+1} \, ,
\end{eqnarray}
where $b_i$ are the annihilation operators of bosons on site i,
$\hat{n}_i = b_i^\dagger b_i^{\phantom{\dagger}}$ the number of
particles on site i, $t$ is the hopping matrix element. $U$ and $V$
are on-site and nearest-neighbor repulsion, and $\mu$ is the chemical
potential.  The energy scale is set by choosing $U = 1$.

The range of the interactions depends on the individual experimental
situation. In general the lattice underlying the system is not an
atomic lattice, but a larger structure like a Josephson-junction or a
grain in a superconductor.  In Josephson-junctions the relevant bosons
can be cooper pairs or fluxes, resulting in different interactions.
 
 As a starting point we first consider the case of on-site repulsion
 only.  In the $(\mu,t)$-plane Mott-insulating regions are surrounded
 by the superfluid phase \cite{Fisher:1989}. These phases are
 separated by two types of phase transitions. On the constant density
 line the transition is driven by phase fluctuations and is of the
 Berezinskii-Kosterlitz-Thouless (BKT) type. The phase transition
 at the sides of the insulator, the generic phase
 transition\cite{Fisher:1989}, is driven by density fluctuations.

The Mott-insulating phases have integer densities and are
incompressible, at the generic phase transition to the compressible
superfluid phase the density of the system changes from commensurate
to incommensurate. The characteristic energy $E_g^{p(h)}$ of this transition
is the energy it costs to create a particle (p) or hole (h)  exitation in
the system.
To calculate this energy we use defect
states \cite{Monien:1996} with the density of the Mott-insulator plus
one additional particle or hole. 
Since the defect states and the insulator
groundstate have fixed densities, a change in the chemical potential by
$\Delta\mu$ does not change the states themselves, but shifts their energy by 
$\Delta\mu N$, where $N$ is the total particle number.
Taking into account that the characteristic energy is zero at the phase 
transition, this gives 
$E_g^{p(h)}(\mu,t) = \mid\mu(t)_c^{p(h)} - \mu \mid ^{z \nu}$, 
where $\mu_c(t)$ is the chemical potential at
the phase transition and the critical exponent $z \nu=1$ \cite{Fisher:1989}.

We use the infinite-size algorithm of the density-matrix
renormalization group (DMRG) \cite{White:1992} with periodic boundary
conditions to determine $E_g^{p(h)}$.  While the maximum number of
particles per site in the Bose-Hubbard model is $n = \infty$, it has
to be cut off for practical calculations with the DMRG. A maximum
occupation number of $n=5$ \cite{Comment2} turned out to be sufficient.

\begin{figure}[t]
  \begin{center}
    \epsfig{file=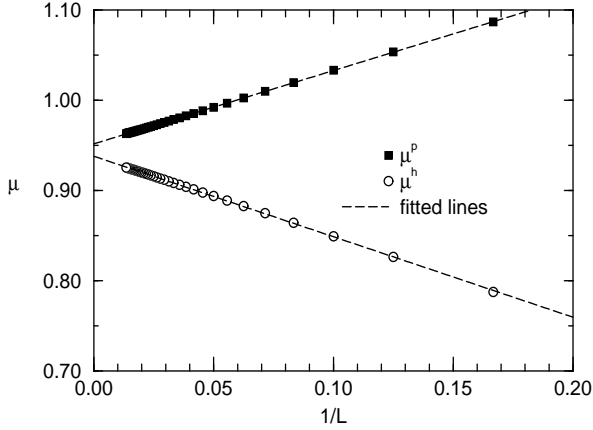, height=\figheight,width=\figwidth, angle=-90}
    \caption[The particle and hole excitation gap]
    { The particle and hole excitation gaps plotted against the
      inverse system size. The straight lines are fitted to find
      $\mu_c$ in the thermodynamic limit. The same scaling behavior has been 
     found in all cases. ($\rho = 1$, $t=0.25$, $U=1$, $V = 0.4$)}
    \label{Gap.extrapolation}
  \end{center}
\end{figure}

Since the two defect states, one with an additional particle, one with
an additional hole, are needed to calculate the energy $E_g^{p(h)}$,
they are used as additional target states in the DMRG. Systems of up
to 76 sites are calculated, keeping $128$ states in each iteration.
The chemical potential $\mu_c^{p(h)}(t) = \mid E_g^{p(h)}(t) - \mu(t)
\mid$ of the phase transition is calculated for various system sizes.
The thermodynamic limit is found by extrapolating to infinite system
size (Fig.\ref{Gap.extrapolation}). Repeating this calculation for various $t$
gives the phase boundaries.

At the tips of the insulator lobes the phase transition is driven by
phase fluctuations.  The characteristic length of this transition is
the correlation length
\begin{equation}
  \label{correlation_length}
  \xi^2 = \sum_r r^2 \Gamma(r) / \sum_r \Gamma(r)
\end{equation}
of the correlation function
\begin{equation}
  \Gamma(r) = \langle b^\dagger(0) b(r) \rangle \;.
\end{equation}
The corresponding energy $E_g$ is the energy gap between the
groundstate and the first excited state with the same density: $E_g
\sim \xi^{-z}$ with the critical dynamical exponent $z = 1$.

The correlation length in the thermodynamic limit can be found by
extrapolating from finite systems: $\xi_L=\xi_\infty + a/L + b \,
\exp{(-L/c)}$, where $L$ is the system size and $a,b,c$ are fitting
constants. The exponential term is small ($a/b>2$), with $c \approx
3$, the results are not changed significantly by neglecting the
exponential term.  The phase transition is of the BKT type,
where the correlation length diverges like
\begin{equation}
  \label{KT_behavior}
  \xi \sim \exp{(\frac{const.}{\sqrt{t_c-t}})} \;.
\end{equation}
One way to find the critical point is fitting (\ref{KT_behavior}) to
the calculated data \cite{Elesin:1994,Kashurnikov:1996.2,Pai:1996}.
But by changing the fitting parameters, this function can go to zero
arbitrarily slow, hence this method is very sensible to numerical
errors and the choice of data points, and we will not use it.

Instead we locate the BKT transition using the analogy of the
superfluid phase to the Luttinger liquid \cite{Haldane:1981}. In the
superfluid phase the correlation function $\Gamma$ decays
algebraically:
\begin{equation}
  \label{parameter.K}
  \Gamma^{SF}(r) \propto r^{-K/2} \,.
\end{equation}

\begin{figure}[t]
  \begin{center}
    \epsfig{file=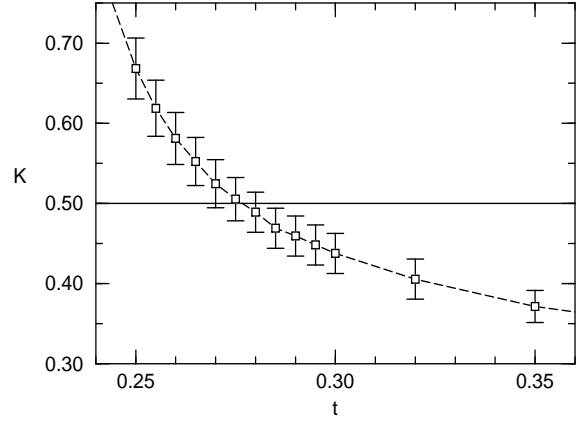, height=\figheight,width=\figwidth,
	 angle=-90}
    \caption
    { $K$ plotted against $t$. 
	 $K_c=1/2$ at $t_c=0.277 \pm 0.1$ ($U =
      1$, $V = 0$). }
    \label{Fig.K_vs_t.MI.noU1}
  \end{center}
\end{figure}

The exponents $K_c$ at the phase transitions are known from Luttinger
liquid theory \cite{Giamarchi:1992.2}. At the BKT transition with
$\rho=1$, $K$ is expected to be $K_c = 1/2$. An algebraic function is
fitted to the correlation functions calculated with DMRG for different
system sizes. Due to the periodic boundary conditions the decay of
$\Gamma$ is very close to algebraic even in small systems.  The
thermodynamic limit of $K$ is found by extrapolating $K(L) = K -
const./L$. Since the decay of $\Gamma^{SF}(r)$ is very close to (\ref{parameter.K}), the main source of errors in $K$ is this extrapolation from finite 
system sizes. We find the phase transition at $t_c = 0.277 \pm 0.01$.

This is in good agreement with $t_c = 0.275 \pm 0.005$ found in an
exact diagonalization approach\cite{Elesin:1994}, and in qualitative
agreement with the Bethe-Ansatz solution $t_c^{BA}=1/(2 \sqrt{3})
\approx 0.289$ for the truncated model with a maximum of $n=2$
particles per site \cite{Krauth:1991}. Three works find somewhat
bigger values $t_c = 0.298$ \cite{Pai:1996}, $t_c = 0.304 \pm
0.002$\cite{Kashurnikov:1996} and $t=0.300\pm0.005$
\cite{Kashurnikov:1996.2}. Early QMC simulations resulted in
$t_c=0.215\pm0.01$ \cite{Batrouni:1992}. The range of these results
demonstrates that determining the location of the BKT transition is 
ill-conditioned.

\begin{figure}[t]
  \begin{center}
    \protect\epsfig{file=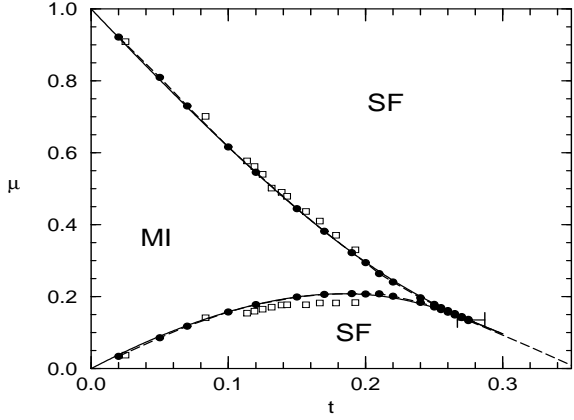,
      height=\figheight,width=\figwidth,angle=-90}
    \protect\caption[]{ The phase
      diagram without nearest-neighbor interaction (MI:Mott-insulator with
      density one, SF:superfluid phase).  The solid lines
      show a Pade analysis of 
      12th order strong coupling expansions \cite{NewExpansion},
      the boxes show Quantum Monte Carlo data \cite{Batrouni:1992}.
      The circles are the DMRG results, the dashed lines indicate the
      area with integer density. The error bars in the $\mu$ direction are 
        smaller than the circles, the error bar in the $t$ direction
         is the error of the BKT transition.
 ($U = 1$, $V = 0$). }
    \label{Fig.phasediagram.MI.noU1}
  \end{center}
\end{figure}

Fig. \ref{Fig.phasediagram.MI.noU1} shows the phasediagram in the  
$(\mu,t)$-plane, including the generic phase boundaries and the location of the BKT transition. For $t\gtrsim 0.225$ we find that the lower phase 
boundary is bending down. 
This means that the Mott-insulator phase is reentrant as a
function of $t$, an unusual feature that has not been observed before.
It implies that increasing $t$, which corresponds to increasing the
kinetic energy, can lead to a {\em reentrance} phase transition from the
superfluid phase to the insulator phase.

In a study inspired by this work a high order strong coupling expansion
\cite{NewExpansion} was used to determine the phase diagram. The phase
boundaries found in that work are in excellent agreement with our DMRG
results, demonstrating the high numerical accuracy that can be achieved
with the DMRG. The strong coupling expansion study confirms the
existence of the reentrant phase transition first found in this work.

One experimental realization of a one dimensional bosonic lattice
system is provided by fluxes in a Josephson-junction array.  In a
recent experiment \cite{Oudenaarden:1996} Mott-insulators with flux
densities $1/3$, $1/2$, $2/3$, $1$.. were found.  In the Bose-Hubbard model
insulators with these densities can only appear in the presence of
longer ranged interactions. A first step in understanding these longer
ranged interactions is studying the effect of repulsion between
nearest neighbors.

In the presence of nearest-neighbor interaction a new insulator phase
appears at half integer densities. It is a charge density wave phase (CDW)
with a wavelength of two sites, and like the Mott-insulator at integer
density it has an excitation gap and is incompressible. The
crystalline order is characterized by the non-zero structure factor
\begin{equation}
 S_\pi=\frac{1}{N^2} \sum_{ij} (-1)^{\mid i-j \mid} 
 \langle \hat{n}_i \hat{n}_{j}\rangle \;.
\end{equation}
$S_\pi$ in the thermodynamic limit is determined by extrapolating from
DMRG calculations for finite systems. A maximum particle number of
four particles per site is chosen for $\rho=\frac{1}{2}$. Since the
groundstate and the first exited state in the CDW
are degenerate in the thermodynamic limit, and close to
degeneracy in finite systems, the first exited state is used as an
additional target state.  $S_\pi$ is found to scale like $S_\pi(L) =
S_\pi + a / L + b \, \exp{(-c/L)}$. The exponential term gives only a
very small contribution ($a/b>10$), $c$ is of the order $10-20$.

In contrast to the transition from the Mott insulator to the
superfluid, which is governed by superfluid order only, the transition
from the CDW to the superfluid is governed by
superfluid and crystalline order.

There are three possible scenarios for this transition: a) There is a
direct phase transition - the vanishing of crystalline order and the
appearance of superfluid order coincide.  b) There is an intermediate
phase with simultaneous superfluid and crystalline order, the
so-called supersolid phase \cite{MeiselBatrouni}.  c) There is an
intermediate {\it normal} phase with neither superfluid nor
crystalline order.

In higher dimensional bosonic systems supersolids exist, but they have
not been observed in one-dimensional systems so far\cite{Niyaz:1994}.
Recently a {\it normal} phase (scenario c) was found in a
numerical study of the one-dimensional Quantum-Phase model
\cite{Baltin}, which is the high density limit of the Bose-Hubbard
model.  This raises the question whether such a {\it normal} phase
also exists in the Bose-Hubbard model.

\begin{figure}[t]
  \begin{center}
    \epsfig{file=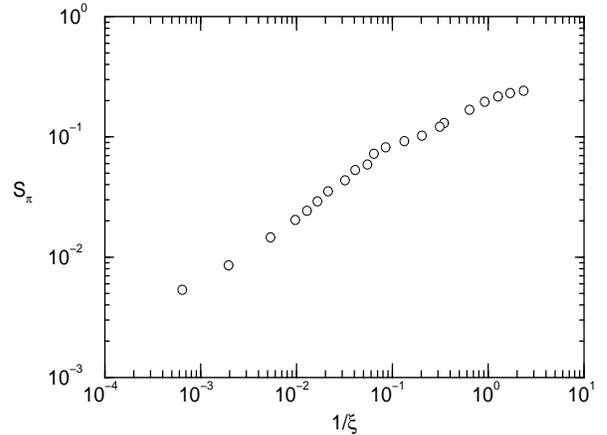, height=\figheight,width=\figwidth,
	angle=-90}
    \caption
    { $S_\pi$ plotted against $\xi$ at the BKT
      transition of the charge density wave phase ($U = 1$ , $V=0.4$).
      The slope is $0.4\pm0.1$.}
    \label{S_vs_xi}
  \end{center}
\end{figure}

The correlation length $\xi$ (\ref{correlation_length}) characterizing
superfluid order diverges in the superfluid phase. If the structure
factor $S_\pi$ is also governed by this correlation length, there is a
direct phase transition from the CDW phase to the
superfluid phase. In that case, close to the phase transition $S_\pi
\sim \xi^{\gamma / \nu} \Phi(\xi/L)$, where $\Phi$ is a scaling
function.  Note that this functional form cannot be transformed to a
power law behavior depending on $t$ due to the BKT behavior of $\xi$
(\ref{KT_behavior}).  To verify the existence of $\Phi$ at the CDW
 to superfluid transition, we plot $S_\pi$ versus $\xi$
(Fig.\ref{S_vs_xi}). We find $S_\pi \sim \xi^{-0.4 \pm 0.1}$, which
shows that there is a direct phase transition and no normal or supersolid 
phase.

The superfluid stiffness in the Luttinger Liquid is always non-zero,
even for large $K$\cite{Fisher:1989}. It has been shown that for $K>1$  
a single weak link or barrier \cite{Fisher_and_Kane,Glazman} becomes relevant, 
reducing the superfluid stiffness to zero by effectively cutting the system 
in two parts.
Since $K=4$ at the sides and $K=2$ at the tip of the
CDW\cite{Giamarchi:1992.2}, this means that the CDW is surrounded by a region
with $K>1$. The normal phase found by Baltin and Wagenblast \cite{Baltin}
was observed in a Quantum Monte Carlo study at finite 
temperatures, where the zero-temperature phase diagram was
extracted with finite-size scaling.  
Baltin and Wagenblast suggested that the normal phase might be the region of
the Luttinger Liquid with $K>1$. But since their calculations were for a
system without impurities, which at incommensurate densities always has a
superfluid stiffness, they should not have observed an effect caused by an 
impurity.
A possible explanation of their result is that thermal fluctuations might
have a similar effect as a weak link. This is supported by the fact that
Baltin and Wagenblast could not determine the scaling function for the 
dependence of the superfluid stiffness on system size and temperature.

\begin{figure}[t]
  \begin{center}
    \epsfig{file=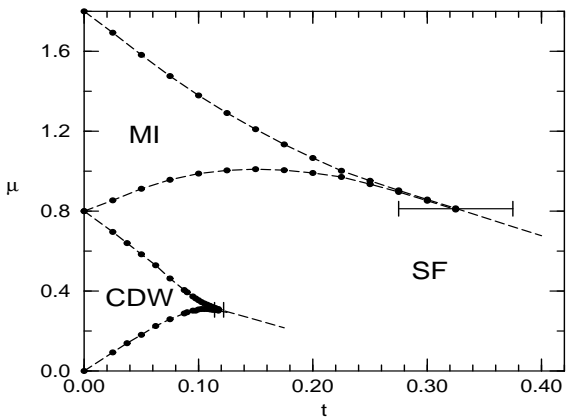,height=\figheight,width=\figwidth,
      angle=-90}
    \caption
    { The phase diagram of the Bose-Hubbard model with
      nearest-neighbor interaction (MI: Mott-insulator with density
      one, CDW:charge density wave with density one half,
      SF:superfluid phase).  The error bars in the $\mu$ direction are
      smaller than the circles, the error bars in the $t$ direction
      indicate the errors of the BKT transitions.  ($U=1$, $V = 0.4
      $).}
    \label{Fig.Phasediagram1}
  \end{center}
\end{figure}

The onset of superfluidity is again determined by the decay of the
correlation functions. At the CDW the critical
exponent is $K_c^{CDW}=2$. The BKT transition is found at
$(t/U)_c^{CDW} =0.118 \pm 0.004$. This is in agreement with $t_c
\approx 0.1$ found with QMC \cite{Niyaz:1994}.  For the Mott-insulator
with density $\rho=1$ the exponent $K$ is found to change very slowly
close to the phase transition, causing a high error margin in our
calculation. We find the critical value $K_c=1/2$ at $t_c \approx
0.325 \pm 0.05$. This indicates that the critical point is shifted to
higher ratios of $t$ by increasing $V$. Within the numerical accuracy
the critical point may also be independent of $V$. This contradicts
QMC results that $t_c$ is reduced if $V$ is increased\cite{Niyaz:1994}.
For $V =0.4$ they found $t_c \approx 0.17$.

Fig. \ref{Fig.Phasediagram1} shows the phase diagram of the
Bose-Hubbard model with nearest-neighbor interaction in one dimension.
To our knowledge this is the first time this phase diagram has been
calculated. The tips of the insulating regions are bending down
towards smaller chemical potentials, which shows the reentrant
behavior already observed in the case without nearest-neighbor
interaction.

In conclusion, we have presented methods to determine the generic as
well as the BKT phase transitions of the Bose-Hubbard model with the
DMRG. At the tips of the insulating regions we found a reentrant phase
transition from the superfluid phase to the insulator.  Including
nearest-neighbor interactions we obtained the new phase diagram and
demonstrated that there is no {\em normal} or supersolid phase, but a
direct phase transition from the CDW to the superfluid phase.

We would like to thank T.~Giamarchi, A.~J.~Millis, R.~Noack,
A.~v.~Otterlo, G.~Sch\"on, H.~Schulz and S.~R.~White for useful and
interesting discussions.

\end{document}